\documentclass[12pt,preprint]{aastex}

\shorttitle{Flux Emergence}
\shortauthors{Galsgaard \etal}

\newcommand{\etal}{{\it et al~}}

\begin{document}

\title{A three-dimensional study of reconnection, current sheets and jets
resulting from magnetic flux emergence in the Sun}

\author{K. Galsgaard\altaffilmark{1}, F. Moreno-Insertis\altaffilmark{2,3},
V. Archontis\altaffilmark{2} and A. Hood\altaffilmark{4}}
\altaffiltext{1}{Niels Bohr Institute for Astronomy, Physics and Geophysics,
  Astronomical Observatory, University of Copenhagen, Juliane Maries vej 30, 2100 Copenhagen {\OE}, Denmark}
\altaffiltext{2}{Instituto de Astrofisica de Canarias (IAC), La Laguna
  (Tenerife), Spain}
\altaffiltext{3}{Department of Astrophysics, Faculty of Physics, Universidad
  de La Laguna, Campus de Anchieta, 38200 La Laguna (Tenerife), Spain}
\altaffiltext{4}{School of Mathematics and Statistics, University of St.
  Andrews,  North Haugh, St. Andrews, KY16 9SS, UK}

\begin{abstract}
We present the results of a set of three-dimensional numerical simulations of
magnetic flux emergence from below the photosphere and into the corona.  The
corona includes a uniform and horizontal magnetic field as a model for a
pre-existing large-scale coronal magnetic system. Cases with different
relative orientations of the upcoming and coronal fields are studied. Upon
contact, a concentrated current sheet with the shape of an arch is
formed at the interface that marks the positions of maximum jump in the
field vector between the two systems. Relative angles above $90^{^\circ}$
yield abundant magnetic reconnection and plasma heating. The reconnection is
seen to be intrinsically three-dimensional in nature
and to be accompanied by marked local
heating. It generates collimated high-speed outflows only a short distance
from the reconnection site and these propagate along the ambient magnetic
field lines as jets.  As a result of the reconnection, magnetic field lines
from the magnetized plasma below the surface end up connecting to coronal 
field lines,
thus causing a profound change in the connectivity of the magnetic regions in
the corona. The experiments presented here yield a number of features
repeatedly observed with the {\sl TRACE} and {\sl YOHKOH} satellites, 
such as the
establishment of connectivity between emergent and pre-existing active
regions, local heating and high-velocity outflows.
\end{abstract}

\keywords{MHD--Sun: activity--Sun: corona--Sun, magnetic fields}

\section{Introduction}
The {\sl TRACE} satellite has provided impressive movies showing the dramatic
phenomena that accompany the emergence of magnetized plasma in the solar
atmosphere \citep{Schrijver99}.  Initially the emerging flux adopts the shape
of a small system of loops in the corona, with their ends anchored in two
localized flux patches in the photosphere.  As the loop system grows in size
it starts feeling the presence of the ambient coronal magnetic field. This
eventually leads to profound changes in the geometry and topology of the
magnetic flux systems, with the emerging field quickly establishing links to
the coronal field.  This process can take place only through magnetic
reconnection that, at the same time, is responsible for significant localized
energy release.
This provides the basic mechanism to create the several million degree plasma
structures observed by the {\sl Yohkoh} soft X-ray telesope above emerging 
flux regions and the cooler loop structures observed by {\sl TRACE} to connect 
emerging regions to the ambient magnetic field.

The first numerical MHD studies of the emergence scenario were
two-dimensional models \citep{Yokoyama+Shibata95,Yokoyama+Shibata96}. These
experiments showed the formation of high-temperature jets, which were
tentatively identified with X-ray jets.  However, the geometry of the complex
structures \citep{Klimchuk97} resulting from flux emergence is intrinsically
three-dimensional.  A number of recent three-dimensional numerical experiments
(\citealt{Fan01,Magara+Longcope01}; \citeyear{Magara+Longcope03};
\citealt{Archontis+ea04, Manchester+ea04}) have studied the rise of buoyant
magnetic flux tubes initially below (but close to) the photosphere into a
non-magnetized corona.  However, the solar corona has a space-filling and
dynamically important magnetic field. Hence, the rise and expansion of newly
emerging magnetic flux cannot proceed unimpeded \citep{Fan+Gibson02}.

Here we present first results of three-dimensional MHD emergence experiments 
that include a
large-scale coronal magnetic field modeled, for simplicity, as a uniform
horizontal field (see \citealt{Archontis+ea04} for details regarding the
MHD equations and the numerical approach).
The experiments reveal a number of strikingly simple
features such as the appearance of an arch-shaped high current density sheet
containing the reconnection sites; change of connectivity of the field lines
that approach the current sheet following regular patterns; formation of
high-velocity plasma outflows emerging from the reconnection region; and plasma
heating associated with the reconnection site. They are discussed in the
following sections.

\section{Model}

\begin{figure}[t]
\epsscale{1.0}
\plotone{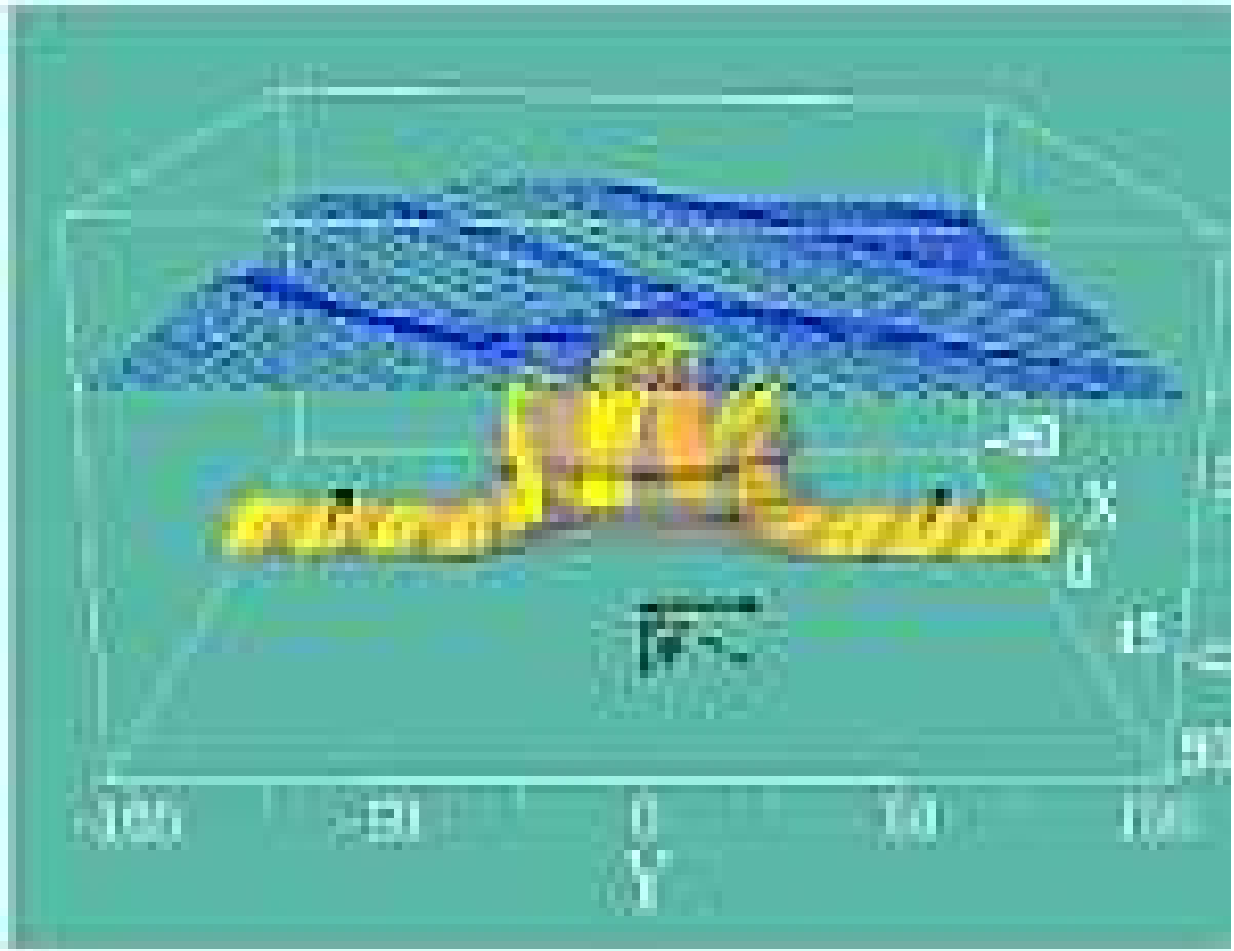}
\caption{\small
Rising flux tube at the time of first contact with the ambient coronal
field, $\phi_0=45^{^{\circ}}$. The transparent isosurface
represents weak magnetic field strength.  Yellow magnetic field lines
wind their way around the tube, while blue field lines shows the structure
of the coronal field. }
\label{fig_plasmahill}
\end{figure}

The initial setup of the experiments follows previous papers \citep[and
references therein]{Fan01,Archontis+ea04}, in which a horizontal, magnetic
flux tube with twisted field lines of constant positive pitch is placed $2$
Mm below the photosphere in a highly stratified environment. The
stratification includes an isothermal photosphere (thickness $1.5$ Mm, at
$T=6500$ K), an isothermal corona (at $T=10^6$ K, with thickness $11.9$ Mm),
a transition region with a steep temperature gradient joining them, and,
at the bottom of the box, an adiabatically stratified region of $3$ Mm
thickness that simulates the uppermost layers of the solar interior.  Unlike
the previous three-dimensional models, an initially uniform and horizontal
magnetic field is
included in the coronal part of the domain with a field strength such that
$\beta \approx 6\thinspace10^{-2}$ and an orientation, relative to the tube
axis, that is used as an input parameter. To initiate the experiment, a
density deficit is incorporated into the tube that is maximum at the center
and goes to zero when moving along the tube axis toward either boundary of
the box. The buoyancy force, therefore, causes the development of a rising
magnetic loop in the central part of the box, which reaches the photosphere
some $t= 12$ minutes after initiation of the experiment. 
Once in the atmosphere, the magnetic pressure drives the further rise of the
magnetized plasma through the transition region and into the corona: the
latter is reached after about $t=25$ minutes.
The rise results in a huge expansion of the plasma: 
the rising loop adopts the shape of a large, somewhat flattened ball, 
or hill, with the field lines showing a fan-like shape
(Fig.~\ref{fig_plasmahill}). On reaching the corona, it bumps against the
coronal field. During the rise phase, the field line windings increase their
radius, and, following simple laws of flux conservation \citep{Parker79}, the
magnetic field at the loop top becomes almost perpendicular to the initial
tube axis (i.e., $\phi \to 180^{^{\circ}}$, with $\phi$ 
the standard polar angle measured in any horizontal plane from the
positive $x$-axis: see Fig.~\ref{fig_plasmahill}).

\section{Results}
When new magnetic flux emerges in the vicinity of existing active regions,
the angle between the emerging and coronal fields can take any value.  To see
the impact of the relative orientation on the field evolution, we have
conducted a series of experiments with the direction of the initial coronal
magnetic field (that we label using its polar angle, $\phi_0$) being changed
from $\phi_0=0$ to $180$ in steps. Seen from the top of the upcoming
plasma, the first case corresponds to the two flux systems being
approximately antiparallel, whereas in the latter they are roughly parallel.
The numerical simulations reveal that the dynamical evolution of the
resulting interaction between the emerging and the coronal flux critically
depends on the relative horizontal angle (180-$\phi_0$) between the two flux
systems at the time of first contact, as explained in the following sections.

\subsection{Arch-like current sheet and heating}\label{sec_current_sheet}

When the emerging flux hits the coronal field, the latter gets pushed upward
and the interface between them becomes dented, thus creating a hill-shaped
structure protruding into the corona.  
In the case when the two fields are aligned ($\phi_0=180^{^{\circ}}$), no
reconnection can initially occur and the emerging field continues to push its
way upward, thus increasing the height of the hill. Initially, the coronal
field is just bent upward with a small sideways expansion. However, it is
energetically favorable for those field lines to slide down the hill so
that they end up passing around it at nearly constant height. The
rearrangement of the coronal field occurs on the local Alfv{\'e}n timescale,
which is much faster than the timescale of rise. 
Only at a later stage does the nonalignment of the
two flux systems become sufficient for some reconnection to occur.

\begin{figure}[t]
\epsscale{1.02}
\plotone{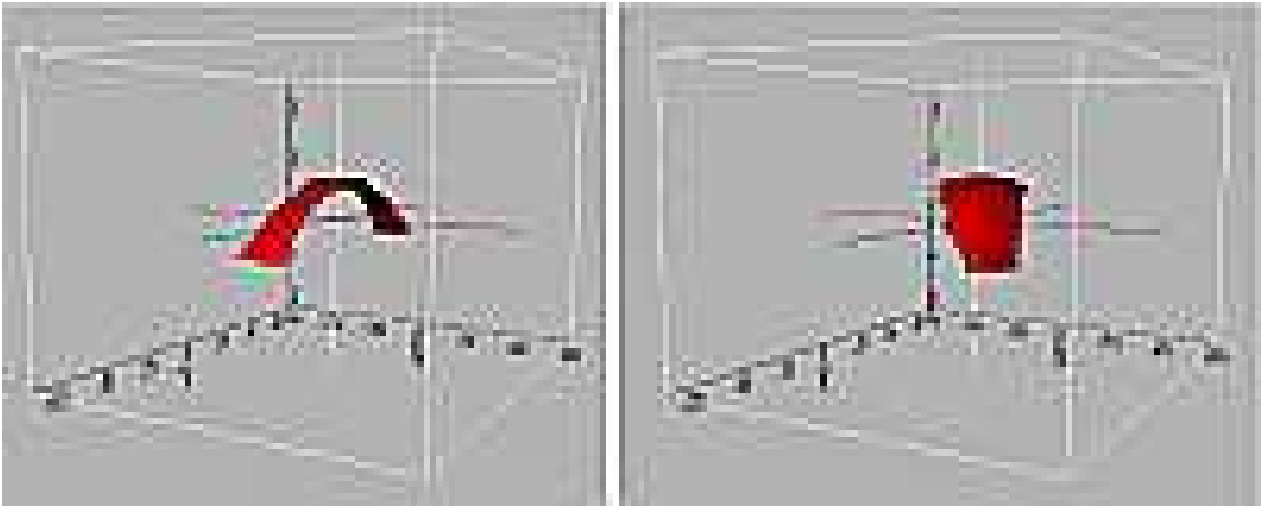}
\caption{\small Arch-shaped current sheet resulting from the contact of the upcoming 
  and ambient flux systems for two cases: antiparallel ($\phi_0=0^{^{\circ}}$, left), 
  and perpendicular ($\phi_0=90^{^{\circ}}$, right). 
  Their vertical mid-planes form an angle with the $z-y$ plane of about $5^{^{\circ}}$
  and $35^{^{\circ}}$, respectively. \label{fig_current_arch} }
\end{figure}

In the opposite limit ($\phi_0=0$), the two flux systems are nearly
antiparallel. Here the discontinuous jump in the magnetic field between
either side of the interface initially yields a dome-shaped current surface.
After a transient evolution, the current surface concentrates into a rather
narrow and curved sheet arching over the summit of the emerging plasma
(Fig.~\ref{fig_current_arch}, {\sl left panel}).  
This current arch is contained in a vertical plane that is rotated some 
$5^{^{\circ}}$
away from the $yz$ plane, an effect of the nonperfect alignment of
field lines of the two systems. 

Going now to cases with an intermediate initial orientation of the ambient
coronal field (i.e., $\phi_0$ between $0$ and $180^{^{\circ}}$), the current sheet
continues to adopt the form of an arch that passes through the
summit point of the emerging hill (fig.~\ref{fig_current_arch}, {\sl right
panel}). Now, however, that the vertical midplane of the arch is rotated with
respect to the $y-z$ plane by an angle that increases  nearly following a 
linear law with $\phi_0$, reaching $35^{^{\circ}}$
for $\phi_0 = 90^{^{\circ}}$ and $50^{^{\circ}}$ for $\phi_0= 135^{^{\circ}}$.
This behavior can
be understood as follows: the nonreconnecting component of the field changes
direction as $\phi_0/2.$ with its magnitude the projection of the field
components onto this direction. Hence, to lowest order approximation, the
magnitude of the reconnecting field components changes as $\cos(\phi_0/2)$.
Given a constant thickness of the current sheet, the electric current inside
it will change in the same manner. These assumptions are generally supported
by the experiments. For example, the peak current density, reached at the
summit of the hill, is seen to follow quite closely a $\cos(\phi_0/2)$ law
(with the peak value reaching $175$ A cm$^{-2}$).

The high current intensity is associated with intense ohmic heating.  
Again, the peak temperature in the current sheet shows a strong dependence
with the initial angle: for $\phi_0 = 0$, the maximum $T$ is $1.1\,10^7$
K, decreasing to $4.6\times10^6$ K for $\phi_0 = 45^{^{\circ}}$ and
$2\times 10^6$ K for $\phi_0 = 90^{^{\circ}}$.

\subsection{The antiparallel case: reconnection and high-velocity
  outflows}\label{sec_antiparallel} 

\begin{figure}[t]
\epsscale{1.0}
\plotone{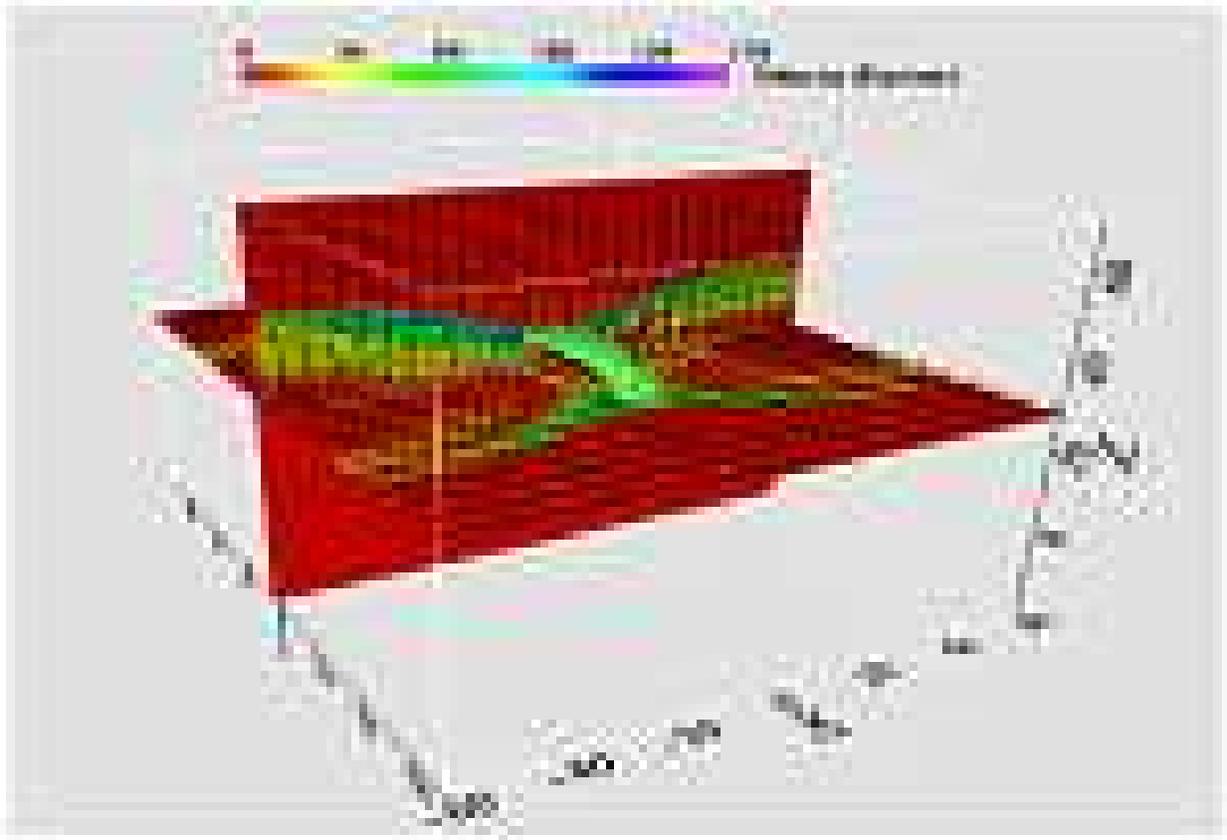}
\caption{\small Three-dimensional view of the current sheet, the velocity magnitude, and
the magnetic field vector for the antiparallel ($\phi_0$=0) case. Shown are a
vertical and a horizontal cut through the domain, each with a color
map of the velocity magnitude and, superposed, the magnetic field
vector. \label{fig_2Dslices}} 
\end{figure}

Let us consider in some detail the physics taking place at the interface
between emerging plasma and ambient field for the antiparallel case (i.e.,
$\phi_0=0$), which 
provides the simplest configuration.  Fig.~\ref{fig_2Dslices} shows a global
three-dimensional view with the help of a vertical and a horizontal 
two-dimensional cut.
The cuts contain a color map of the modulus of the velocity, as well as the full
magnetic vector. For clarity, one-half of the arched current sheet is also
shown (colored green).  Reconnection is occurring all along the current arch,
with field lines of opposite sense being pressed together from above and
below, and the reconnected field lines going away sideways from the site.  
In a constant $x-z$ plane close
to the summit point of the plasma hill, the magnetic structure is comparable
to a bent two-dimensional $x$-point that has collapsed into two $y$-points connected by a
current sheet through which reconnection proceeds \citep{Syrovatski71}.

The effects of reconnection are particularly apparent at the edges of the
current sheet in both the horizontal and vertical cuts: here, the reconnected
field lines detach from the current sheet and are ejected sideways,
with the associated launching of hot high-velocity plasma jets.  
At only a short distance from the current sheet
(e.g., several hundred kilometers in the figure), the plasma flows almost along the
coronal magnetic field lines. High velocities, in excess of $100$ km
s$^{-1}$, are reached both close to the reconnection site as well as in the
higher part of the jet, with the peak velocities typically reaching the local
Alfv{\'e}n velocity. Completing the 3D image of the outflows one can see that the
high-speed outflow volumes have the shape of two curved caps, one on each side
of the current arch, with their crest becoming asymptotically horizontal.

\subsection{The general case: global configuration and outflows}
\label{sec_slanted} 

\begin{figure}[t]
\epsscale{1.0}
\plotone{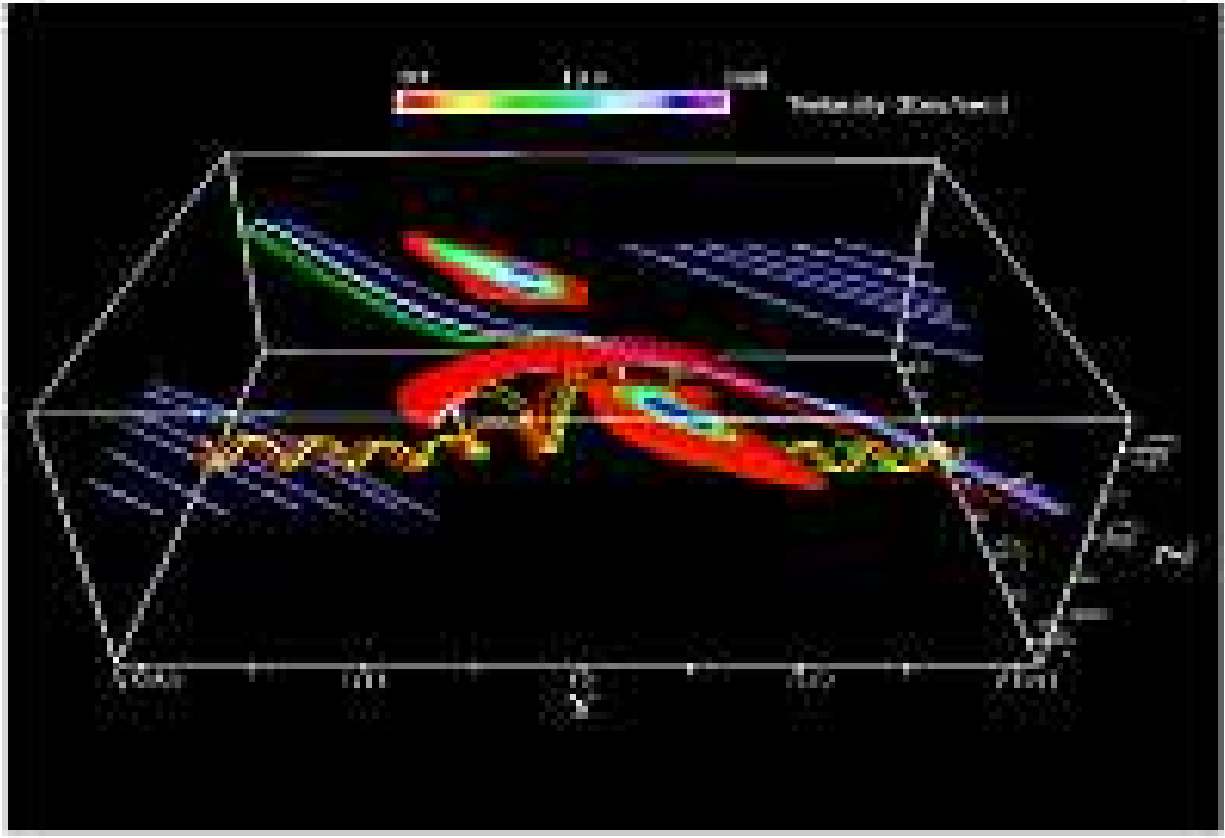}
\caption{\small Location and effect of the reconnection taking place at a 
later time than shown in Fig.1. The image shows the current sheet as an
arch-shaped red isosurface; isosurfaces of the two high-speed jets and
various sets of magnetic field lines \label{fig_composite}}
\end{figure}

The relative simplicity of the structure described in the previous section
and glimpsed in Fig.~\ref{fig_2Dslices} is lost when considering a case with
an arbitrary orientation of the initial coronal field. As a representative
example we take $\phi_0=45^{^{\circ}}$, which
still allows for abundant electric current density, heating and
reconnection. Fig.~\ref{fig_composite} shows the global configuration and the
relative position of the current sheet ({\sl red arching stripe}), 
outflow jets,
ambient coronal field ({\sl thin straight blue lines}), 
and the magnetic tube still buried below the
photosphere ({\sl spiraling yellow field lines}).
The vertical midplane of the current sheet subtends
in this case an angle of approximately $20^{^{\circ}}$ deg with the $y-z$
plane. The
field lines shown belong to four different classes: (1) blue coronal field
lines, (2) yellow flux tube field lines, and (3) and (4) red and green reconnected
field lines connecting the flux tube to the coronal field.  Reconnection is
occurring all along the current sheet, and as before, high-velocity outflows
are being launched away from the sheet. 
Figure~\ref{fig_composite} shows the location of maximum velocity of
those jets, using isosurfaces of the velocity magnitude: they are visible as
two curved caps that point away from the current sheet close to its
upper level. Velocities of close to $200$ km s$^{-1}$ are reached.
All along the jets there is only a very limited reduction
in velocity, indicating that reconnection jets will be seen over very large
distances in the corona.

\subsection{Reconnection in the general case: 2D vs 3D reconnection}
\label{sec_2d_vs_3d}

Reconnection in the general case has the distinctive feature (when compared
with the antiparallel case; Sec.~\ref{sec_antiparallel}) of being essentially
three-dimensional in nature. For example, two of the sets of coronal field
lines in Fig.~\ref{fig_composite} ({\sl red and green}) traverse the current sheet
at its summit and connect the coronal magnetic system with the magnetic flux
tube: notice the clear nonalignment of the coronal and tube magnetic field
on either side of the current sheet.  Because of the nonzero resistivity, the
orientation of the field changes smoothly through the current sheet, so that
around the summit point of the arch, the field line system resembles a
smoothed rotational discontinuity.  The magnetic pressure decreases toward
the center of the sheet, with the magnetic field here containing only the
nonreconnecting part of the initial magnetic field vectors. Pressure balance is
achieved via a gas pressure maximum at the center of the sheet.
In a general 3D reconnection situation,
there is a finite noneconnecting component at the center of the sheet, and
the gas pressure maximum is not too pronounced. Toward the flanks of the
sheet, however, the nonreconnecting component is small and the magnetic
pressure can become $10^3$ times smaller than the ambient magnetic pressure.

\begin{figure}[t]
\epsscale{1.}
\plotone{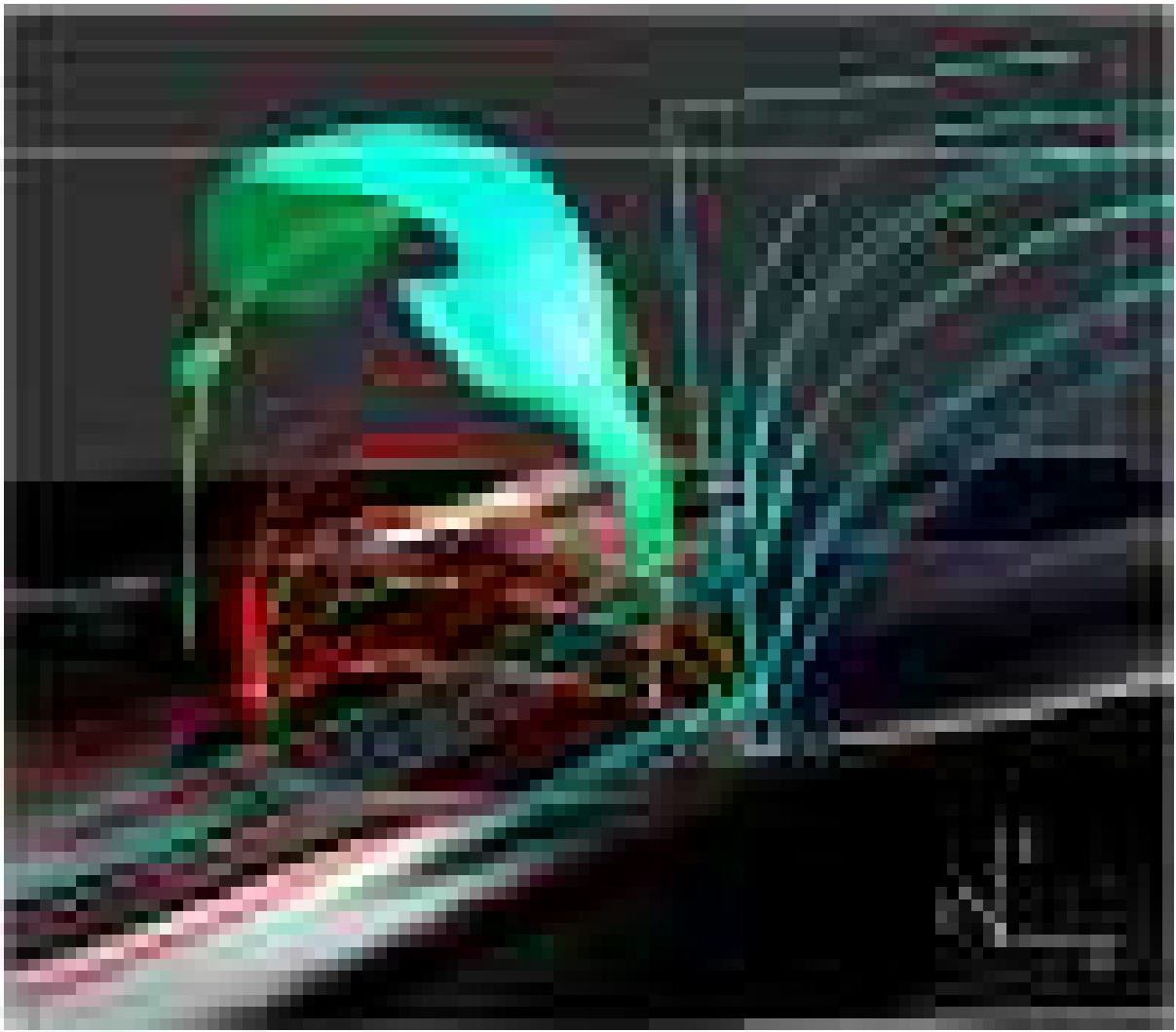}
\caption{\small Close-up of the magnetic field line topology around one end
of the current sheet seen in Fig.~\ref{fig_composite}.  
The shaded image at the bottom is an intensity map of
$v^2$. \label{fig_reconnection_site}}

\end{figure}

The field topology can be illustrated by drawing field lines close to the
lateral end of the current sheet (Fig.~\ref{fig_reconnection_site}). Four
distinct types are shown together with the full current sheet. 
Both the red and dark blue field lines are the direct
result of reconnection between the emerging and coronal fields. They
highlight the asymmetry between the two sides of the reconnection site: the
red field lines are highly curved close to the reconnection point and, hence,
will generate a faster outflow than the less curved dark blue field lines. The
yellow field lines belong to the emerging flux system within the plasma hill,
and have not yet undergone any reconnection. The light blue field lines, finally,
are part of the ambient coronal system but pass close to the reconnection
site.  The traditional reconnection picture is linked with the concept that
the magnetic field lines break near the center of the current sheet. As seen
here, this need not happen for less symmetric situations than assumed in the
simple 2D scenarios. In fact, in our experiments the flanks of the current
sheet are maintained by the rapid change in orientation of the red
reconnected field lines.  Moreover, we consistently find the reconnection
point located far out to one side of the current sheet, all of which
indicates a departure from simple two-dimensional properties.

\section{Discussion}

The present experiments show that interaction between the emerging flux and
the existing coronal magnetic field is unavoidable and that it can cause a
highly dynamical evolution. They also show that the efficiency of the
interaction between the two flux systems critically depends on the relative
orientation between their magnetic field components. Fast magnetic
reconnection can be delayed when their orientations are nearly parallel,
while it is easily initiated when they are nearly antiparallel. An
arch-shaped current sheet is formed; all along its rim the reconnection
generates high-velocity outflows, which are seen propagating horizontally
away from the emergence site in the ambient coronal field. The experiments
also show that nonsymmetric features arise when the field vector of the two
flux systems is not contained in the same vertical plane. Furthermore it is
found that simple extensions of the magnetic reconnection process in
traditional two-dimensional models cannot accomodate the fully
three-dimensional features occurring in the reconnection in these experiments.

Scaling the model to solar conditions, the experiment covers a horizontal
domain of $20$ x $23$ Mm$^{2}$, corresponding to the size of a small active
region. The emergence phase into the corona covers only about 30
minutes. Nonetheless, the reconnection jets obtained in the most favorable
cases have high velocities, with peak values of the order of $400$ Km s$^{-1}$.
At the same time, the magnetic dissipation heats the plasma significantly in the
current sheet, with temperatures in the antiparallel case reaching close to
$10^7$ K. 
Higher numerical resolution - lower effective resistivity - will
increase the energy dissipation in the current sheet and with that the peak
temperature.  This will be countered by including anisotropic
heat conduction and optically thin radiation in the energy equation - the
task for a future investigation. 

The results obtained in our experiments compare favorably with
observations, where typical jet velocities \citep{Shimojo+ea96} are between
$10$ and $1000$ km s$^{-1}$, with an average velocity $200$ km s$^{-1}$ and 
where plasma
temperatures of many million degrees are often reported in connection with
fast energy release events, such as flares.

\acknowledgments 

Computational time on the Linux clusters in St. Andrews
(PPARC and SRIF funded) and at the Instituto de Astrofisica de Canarias
(partially funded through the MCYT) as well as financing through the
European Comission's PLATON Network (HPRN-CT-2000-00153) and project
AYA2001-1649 of the Spanish MCYT are gratefully acknowledged.  K.G. was
supported by the Carlsberg Foundation in the form of a fellowship.

\clearpage
\clearpage

\clearpage

\clearpage

\clearpage

\end{document}